\ifdefined\isArxiv
\documentclass[lettersize,journal]{IEEEtran}
\else
\documentclass[12pt, draftclsnofoot, onecolumn,lettersize]{IEEEtran}
\fi
\ifCLASSINFOpdf
\else
\fi
\usepackage{amsmath}
\usepackage{multicol}
\usepackage{mathtools}
\usepackage{dsfont}
\usepackage{subcaption}
\usepackage{booktabs}
\usepackage{hyperref}

\usepackage[dvipsnames]{xcolor}
\usepackage{soul}
\sethlcolor{yellow}

\usepackage{xcolor}

\usepackage{wrapfig}

\begin{document}
\title{Multi Digit Ising Mapping for Low Precision Ising Solvers}

\author{\IEEEauthorblockN{Abhishek~Kumar~Singh, Kyle~Jamieson}\\
\IEEEauthorblockA{\textit{Department of Computer Science, Princeton University}}}

\maketitle

\begin{abstract}
The last couple of years have seen an ever-increasing interest in using different Ising solvers, like Quantum annealers, Coherent Ising machines, and Oscillator-based Ising machines, for solving tough computational problems in various domains. Although the simulations predict massive performance improvements for several tough computational problems, the real implementations of the Ising solvers tend to have limited precision, which can cause significant performance deterioration. This paper presents a novel methodology for mapping the problem on the Ising solvers to artificially increase the effective precision. We further evaluate our method for the Multiple-Input-Multiple-Output signal detection problem.
\end{abstract}

\section{Introduction}
\label{intro}
Physics-inspired computation and Ising solvers provide a radically different alternative to conventional computation. Researchers have extensively studied Ising machines, like Quantum annealers, Coherent-Ising machines, \textit{etc.}, for solving tough computational problems across several domains, and have shown promising performance gains. It has been demonstrated that Ising solvers can tackle abstract NP-Hard problems, like MAX-CUT~\cite{isingMaxCut} and SAT~\cite{qaSAT_MAXSAT}, as well as several real-life problems like stock market prediction and Multiple-Input-Multiple-Output~(MIMO) signal detection~\cite{di-mimo,ri-mimo,minsungParallelTemp,minsung2019,kim2022warm,pSuccessQA}. However, unlike software-based Ising solvers and simulation models that utilize floating-point precision, real implementations tend to have limited precision for representing problem coefficients. These precision limitations can severely impact the gains predicted by simulated models. However, real implementations~\cite{cobi} are often limited by hardware, and providing high precision for representing the problem coefficients remains a significant challenge. In this paper, we present a novel Ising mapping that can artificially increase the effective precision of an Ising solver, allowing it to achieve arbitrarily high precision without any change in the underlying hardware.

MIMO signal detection presents a unique challenge for Ising solvers by imposing extremely aggressive timing requirements. Practical systems like LTE require solving thousands of problems within a few milliseconds~\cite{ri-mimo}. Researchers have proposed several methodologies for utilizing Ising solvers for MIMO detection and have shown promising performance improvements via software simulations~\cite{di-mimo,ri-mimo,minsungParallelTemp, minsung2019,kim2022warm,pSuccessQA}. In this paper, we demonstrate the performance of MIMO detection on the real implementation of an Ising solver and demonstrate that the performance that our proposed methodology can mitigate the performance loss due to the low precision of the underlying hardware.

The rest of the paper is organized as follows. Section~\ref{sec:primer} provides a primer on Ising optimization and the MIMO detection problem. Section~\ref{sec:design} describes our proposed methodology and Section~\ref{sec:eval} presents the evaluation of our proposed methods for different MIMO scenarios. 

\section{Primer}
\label{sec:primer}
\subsection{Ising Optimization}
The Ising optimization problem~\cite{di-mimo} is an NP-Hard~\cite{isingNpHard1}, quadratic binary optimization problem given by,
\begin{equation}
    \arg \min  -\sum_{i \neq j}J_{ij}s_{i}s_{j},
\end{equation}
where $h_i$, $J_{ij}$ are real-valued problem coefficients and $s_i$ and $s_j$ are spin variables that take values $\{-1,1\}$. 

An Ising solver is an algorithm/hardware that takes the coefficients of the Ising optimization problem as an input and outputs a candidate solution. The candidate solution can be the ground state of the problem or a local minima. A few examples of such Ising solvers are Quantum Annealing~\cite{hauke2020perspectives}, Coherent Ising machines~\cite{wang2013coherent,marandi2014network,mcmahon2016fully,inagaki2016coherent,dopo,oeo}, digital-circuit Ising solvers~\cite{yamaoka201520k,aramon2019physics,goto2019combinatorial,goto2021high,leleu2020chaotic}, bifurcation machines~\cite{tatsumura2021scaling}, photonic Ising machines besides CIMs~\cite{roques2020heuristic,prabhu2020accelerating,babaeian2019single,pierangeli2019large}, spintronic and memristor Ising machines~\cite{sutton2017intrinsic,grollier2020neuromorphic,cai2020power}, and oscillation-based Ising machines~\cite{oim}. Borrowing the terminology from Quantum annealing literature, we refer to each run of the Ising solver as an \textit{anneal}. It is common practice to solve the same problem multiple times on the Ising solver and generate several candidate solutions, \textit{i.e.}, to perform multiple anneals. Each anneal generates a candidate solution and finally, the best solution is selected as the output. 

\subsection{MIMO Detection problem}
A MIMO system~\cite{di-mimo} consists of a receiver node with $N_r$ receiver antenna and one or more transmitters with a total of $N_t$ transmit antennas. The signal transmitted from the transmitters can be described by a complex-valued $N_t \times 1$ transmit vector $\mathbf{x}$, and the signal at the receiver can be described by a complex-valued $N_r \times 1$ receive vector $\mathbf{y}$. The channel between the transmitters and the receivers is represented via a complex-valued $N_r \times N_t$ channel matrix $\mathbf{H}$, and the input-output relationship is given by
\begin{equation}
    \mathbf{y} = \mathbf{H}\mathbf{x} + \mathbf{n},
\end{equation}
where $\mathbf{n}$ is the $N_r \times 1$ noise vector. Each element of the transmit vector $\mathbf{x}$ is derived from a discrete complex-valued set $\Omega$, \textit{i.e.}, $\mathbf{x} \in \Omega^{N_t}$. Let us assume that the channel $\mathbf{H}$ is known at the receiver, then the goal of MIMO detection is to find $\mathbf{x}$ given $\mathbf{y}$ and $\mathbf{H}$. Under the assumption that the system has Gaussian white noise, the optimal detection method is the Maximum Likelihood detector~(MLD):
\begin{equation}
    \hat{\mathbf{x}} = \arg \min_{\mathbf{u} \in \Omega^{N_t}} ||\mathbf{y} - \mathbf{H}\mathbf{u}||^2.
\end{equation}

MLD is known to be NP-Hard and is therefore infeasible for practical systems with a large number of transmitters and receivers. It has been shown that using Ising machines to solve the MLD problem can provide significant performance gains and can potentially meet the requirements of a practical system. However, these results were based on software simulations of the dynamics on a Coherent Ising machine and used floating point precision. In this paper, we will evaluate DI-MIMO~\cite{di-mimo}, known to have the best performance for MLD, in precision-limited scenarios, and show that there is significant performance deterioration. 

\section{Design}
\label{sec:design}
In this section, we present our proposed Multi-digit Ising representations which can be used to increase the effective precision of an Ising solver. 

Let us assume that the couplings on the Ising solver can take integer values in the range $[-2C_{max},2C_{max}]$. Since the coupling between spin $i$ and $j$ is given by $K_{ij} + K_{ji}$, without loss of generality, we can impose a constraint that $-C_{max} \leq K_{ij}, K_{ji} \leq C_{max}$ and $K_{ij}, K_{ji} \in \mathcal{Z}$. Therefore, the magnitude of each coefficient can be at most $C_{max}$, leading to a total of $2C_{max}+1$ possible couplings (approximately $log_2(C_{max})$ bits of precision).

To increase the effective precision, instead of quantizing the problem coefficients to integers in $[-C_{max}, C_{max}]$, we quantize them to 3-digit base-q integers, where $q \leq (C_{max} + 1)$. The largest 3-digit base-q integer is given by $M_q = (q-1)(q^2 + q + 1)$, and therefore the Ising coefficients are quantized to integers in $[-M_q,M_q]$. Therefore,
\begin{equation}
     K'_{ij} = \left\lceil \dfrac{M_q(J_{ij} + J_{ji})}{2} \right\rceil,  K'_{ji} = \left\lfloor \dfrac{M_q(J_{ij} + J_{ji})}{2} \right\rfloor.
\end{equation}
Then using the octal representation of $|K'_{ij}| = (q^2e_{ij} + qf_{ij} + g_{ij})$ we can express
\begin{equation}
    K'_{ij} = sign(J_{ij})\cdot|J_{ij}| = sign(J_{ij}) (q^2e_{ij} + qf_{ij} + g_{ij}),
\end{equation}
where $e_{ij},f_{ij},g_{ij}$ take integer values in $[0,q-1]$. If $a_{ij} = sign(K'_{ij})\cdot e_{ij}$, $b_{ij} = sign(K'_{ij})\cdot f_{ij}$, $c_{ij} = sign(K'_{ij})\cdot g_{ij}$, then the corresponding Ising term is given by
\begin{equation}
-q^2a_{ij}s_{i}s_{j} - qb_{ij}s_{i}s_{j} - c_{ij}s_is_j = -a_{ij}(qs_i)(qs_j) - b_{ij}(qs_i)s_j - c_{ij}s_is_j
\end{equation}
where $a_{ij},b_{ij},c_{ij}$ take integer values in $[-(q-1),(q-1)]$. For each spin variable $s_i$, let us create q other spin variables $s_{i,1},s_{i,2},...s_{i,q}$ which will be forced to take the same value as $s_i$, then $qs_i = \sum_{k=1}^qs_{i,k}$. Therefore we can express the Ising term as
\begin{equation}
    -a_{ij}(\sum_{k = 1}^qs_{i,k})(\sum_{k = 1}^qs_{j,k}) - b_{ij}(\sum_{k = 1}^qs_{i,k})s_j - c_{ij}s_is_j
\end{equation}

To force $s_{i,k}$ to take the same value as $s_i$ we add the term $-C_{max}s_is_{i,k}$ to the Hamiltonian. To ensure stronger couplings between the copies of the same spin $s_i$ additional terms of the form $-C_{max}\sum_{k = 1}^q\sum_{l = 1}^qs_{i,k}s_{i,l}$ can be added to the hamiltonian as well.
Therefore, the final Hamiltonian is given by

\begin{equation}
   \arg\min \left[\left(-\sum_{i\neq j}[a_{ij}(\sum_{k = 1}^qs_{i,k})(\sum_{k = 1}^qs_{j,k}) + b_{ij}(\sum_{k = 1}^qs_{i,k})s_j + c_{ij}s_is_j]\right) + \sum_{i}C(i)\right],
\end{equation}
where
\begin{equation}
   C(i) = - C_{max}\sum_{k = 1}^qs_is_{i,k} -C_{max}\sum_{k = 1}^q\sum_{l = 1}^qs_{i,k}s_{i,l}
\end{equation}
represents the additional terms added to make $s_i, s_{i,1}, s_{i,2},...s_{i,q}$ take the same value. 
Note that, each coefficient in the Hamiltonian takes integer values in $[-(q-1),(q-1)]$ where $q \leq C_{max}$. Therefore each coefficient takes integer values in $[-C_{max},C_{max}]$ but the effective precision for the coefficients of the original problem is $log_2(M_q) + 1$ bits (integer values in the range $[-M_q, M_q]$). 

\subsection{2-digit simplification}
In the last section, we used a formulation with three base-q digits to represent the problem coefficients. However, the 3-digit representation requires us to create q copies of each spin variable, increasing the problem size q folds. For Ising solvers, that don't have enough spins to support 3-digit representations, 2-digit representations can be utilized. For a 2-digit representation, $M_q = (q-1)(q+1)$. In a similar way to 3-digit representation, we express the Ising coefficient  $K'_{ij}$ using two base-q digits: 
\begin{equation}
    K'_{ij} = qf_{ij} + g_{ij},
    \label{eq:2digit1}
\end{equation}
which can be then equivalently written as
\begin{equation}
    K'_{ij} = (q+1)f_{ij} + g_{ij}-f_{ij},
    \label{eq:2digit2}
\end{equation}
and the corresponding Ising term (following Eq.~\ref{eq:2digit1} and Eq.~\ref{eq:2digit2}) is given by
\begin{equation}
    -qf_{ij}s_is_j - g_{ij}s_{i}s_{j} = -(q+1)f_{ij}s_is_j - (g_{ij}-f_{ij})s_{i}s_{j}
\end{equation}

Note that, since both $f_{ij}$ and $g_{ij}$ have the same sign ($=sign(J_{ij})$), if $|f_{ij}|,|g_{ij}| \leq C_{max}$, then $|g_{ij}-f_{ij}| \leq C_{max}$. Therefore, either of the two representations can be used to map the problem on the Ising solver. Unlike three-digit representation, instead of creating q copies of each spin variable for the term $qs_i$, we can express the Ising term (following Eq.~\ref{eq:2digit1}) as 
\begin{equation}
    -\alpha^{(1)}_{ij}(\beta^{(1)}_{ij}s_i)(\gamma^{(1)}_{ij}s_j) - g_{ij}s_{i}s_{j}
    \label{eq:factor1}
\end{equation} 
or equivalently (following Eq.~\ref{eq:2digit2}) as
\begin{equation}
    -\alpha^{(2)}_{ij}(\beta^{(2)}_{ij}s_i)(\gamma^{(2)}_{ij}s_j) - (f_{ij}-  g_{ij})s_{i}s_{j}
    \label{eq:factor2}
\end{equation} 
such that $\alpha^{(1)}_{ij}\beta^{(1)}_{ij}\gamma^{(1)}_{ij} = qf_{ij}$ and $\alpha^{(2)}_{ij}\beta^{(2)}_{ij}\gamma^{(2)}_{ij} = (q+1)f_{ij}$. 

Note that, in the first expression (following Eq.~\ref{eq:factor1}), to map the Ising term to the Ising solver, we would need to create $\beta^{(1)}_{ij}$ copies of $s_i$ and $\gamma^{(1)}_{ij}$ copies of $s_j$, and $\alpha^{(1)}_{ij}$ will become the coupling strength. Therefore, to minimize the number of copies needed in the first formulation, we choose $\alpha^{(1)}_{ij},\beta^{(1)}_{ij},\gamma^{(1)}{ij}$ such that: we minimize the $max(\beta^{(1)}_{ij},\gamma^{(1)}_{ij})$, while $|\alpha^{(1)}_{ij}| \leq C_{max}$ and $\alpha^{(1)}_{ij}\beta^{(1)}_{ij}\gamma^{(1)}_{ij} = qf_{ij}$. 

Similarly, in the second formulation (following Eq.~\ref{eq:factor2}), to map the Ising term to the Ising solver, we would need to create $\beta^{(2)}_{ij}$ copies of $s_i$ and $\gamma^{(2)}_{ij}$ copies of $s_j$, and $\alpha^{(2)}_{ij}$ will become the coupling strength. Therefore, to minimize the number of copies needed in the second formulation, we choose $\alpha^{(2)}_{ij},\beta^{(2)}_{ij},\gamma^{(2)}{ij}$ such that: we minimize the $max(\beta^{(2)}_{ij},\gamma^{(2)}_{ij})$, while $|\alpha^{(2)}_{ij}| \leq C_{max}$ and $\alpha^{(2)}_{ij}\beta^{(2)}_{ij}\gamma^{(2)}_{ij} = (q+1)f_{ij}$. Finally, we choose the formulation, that requires to the least number of copies.

While we describe 2-digit and 3-digit representations, it is straightforward to use the same ideas for representations with more digits and achieve arbitrarily high precision. 
\begin{figure}[h]
  \centering
  \includegraphics[width=1\textwidth]{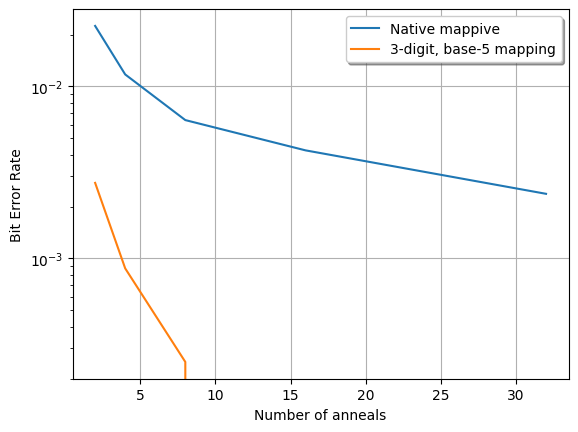}
  \caption{Bit error rate performance of native mapping vs 3-digit base-8 mapping for a $N_t=2$, $N_r =2$ MIMO system with 16-QAM modulation and no noise. We see that the proposed mapping can significantly improve the BER performance.} %like QuAMax~\cite{minsung}.}
  \label{fig:ber2x2_3db5}
\end{figure}
\begin{figure}[h]
  \centering
  \includegraphics[width=1\textwidth]{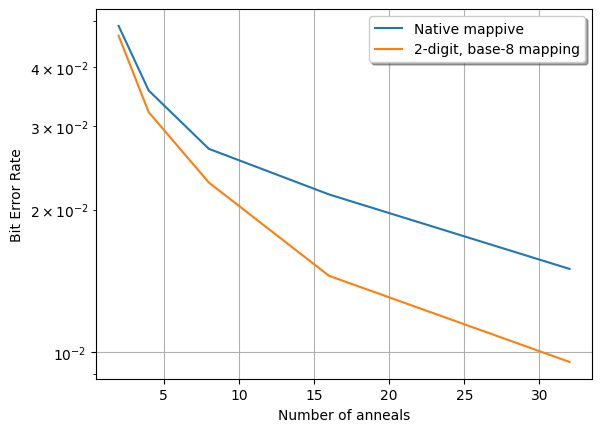}
  \caption{Bit error rate performance of native mapping vs 2-digit base-8 mapping for a $N_t=3$, $N_r =3$ MIMO system with 16-QAM modulation and no noise. We see that the proposed mapping can significantly improve the BER performance.} %like QuAMax~\cite{minsung}.}
  \label{fig:ber3x3_2db8}
\end{figure}

\section{Evaluation}
\label{sec:eval}
In this section, we evaluate the performance of our methods for the MIMO detection problem. We implement the Delta Ising MIMO~(DI-MIMO)~\cite{di-mimo} algorithm, and evaluate its performance on the COBI Ising solver. 
\subsection{COBI Ising solver}
In this paper, we use the COBI Ising solver~\cite{cobi} chip which utilizes a network of coupled oscillators to solve the Ising problems. It can solve the Ising problem given by
\begin{equation}
    \arg \min  -\sum_{i \neq j}K_{ij}s_{i}s_{j},
\end{equation}
where $K_{ij} + {K_{ji}}$  represents the coupling between spin $i$ and $j$. COBI chip implements 29 discrete coupling levels between two spin variables (integer values in [-14,14]), \textit{i.e.}, $\forall i,j$ $|K_{ij} + K_{ji}| <= 14$ and $J_{ij} \in \mathcal{Z}$ (where $\mathcal{Z}$ is the set of all integers).

The default approach for solving an Ising problem on the COBI chip involves scaling the Ising coefficients and rounding to the nearest integer value allowed by the chip. If we assume that the original problem coefficients $J_{ij}$ are scaled such that $|J_{ij}| \leq 1$ then 
\begin{equation}
    K_{ij} = \left\lceil \dfrac{7(J_{ij} + J_{ji})}{2} \right\rceil,  K_{ji} = \left\lfloor \dfrac{7(J_{ij} + J_{ji})}{2} \right\rfloor
\end{equation}
This introduces a quantization error in the mapping, leading to performance deterioration. Since the COBI chip allows a maximum of 59 spin problems, we select the parameters of the mapping such that it can fit the chip. 

\subsection{Bit Error Rate Performance}
First, let us look at a MIMO scenario with $N_t = 2$ and $N_r = 2$, 16-QAM modulation and no noise. We use a 2-digit, base-5 mapping which increases the effective precision to integers in $[-124,124]$, and in Fig~\ref{fig:ber2x2_3db5} we see that the bit error rate performance is drastically better than the native mapping. 

Next, let us look at a MIMO scenario with $N_t = 3$ and $N_r = 3$, 16-QAM modulation and no noise. We use a 2-digit, base-8 mapping which increases the effective precision to integer in $[-63,63]$, and in Fig~\ref{fig:ber3x3_2db8} we see that even with a 2-digit representation the bit error rate performance is significantly better than the native mapping.

\bibliographystyle{IEEEtran}

\bibliography{IEEEabrv,sn-bibliography}
\end{document}